# Exploring Damping Properties of IRIS Bright Points using Deep Learning Techniques

## E. Tavabi[1] & R. Sadeghi[1]

[1]Physics Department, Payame Noor University, Tehran, Iran, 19395-3697

**Abstract.** This study analyzed the Doppler shift in the solar spectrum using the Interface Region Imaging Spectrograph (*IRIS*). Two types of oscillations were investigated: long period damp and short period damp. The researchers observed periodic perturbations in the Doppler velocity oscillations of bright points (BPs) in the chromosphere and transition region (TR). Deep learning techniques were used to examine the statistical properties of damping in different solar regions. The results showed variations in damping rates, with higher damping in coronal hole areas. The study provided insights into the damping behavior of BPs and contributed to our understanding of energy dissipation processes in the solar chromosphere and TR.

---

## 1. Introduction

Solar atmospheric oscillations, particularly Doppler shift oscillations, provide valuable insights into the dynamics and physical properties of the Sun. The network and internetwork regions are of particular interest due to their distinct magnetic field characteristics. However, analyzing these oscillations is challenging due to noise and instrumental effects. Recent studies have revealed oscillatory behavior in coronal BPs, and this study aims to analyze longitu- dinal oscillations in chromospheric BPs using Doppler maps. Deep learning algorithms are employed to overcome the challenges and uncover the underlying oscillatory signals. The study seeks to enhance our understanding of solar atmospheric oscillations and explore their connection with phenomena like spicular mass ejections and transient heating of plasma. The application of deep machine learning holds promise for accurately identifying and character- izing short-period oscillations in network and internetwork regions. By leveraging advanced data processing methodologies, researchers can gain novel insights into the physical mecha- nisms driving solar atmospheric oscillations(Ajabshirizadeh et al. 2011; Chandrashekhar and Sarkar 2015; Krijger et al. 2001; Tavabi et al. 2011; Rojas-Quesada 2021; Sadeghi and Tavabi 2022b; Tavabi et al. 2015; Zeighami et al. 2020; Tavabi 2018; Tavabi et al. 2022).

## 2. Observation

This study utilized data from the *IRIS*, including slit jaw images (SJIs) and spectral data covering a wide range of wavelengths. The data were collected from different solar regions, including equatorial coronal holes, quiet Sun, and active regions. The Mg II h and k lines were specifically studied to understand plasma at low temperatures. By integrating data from multiple sources, researchers gained a comprehensive understanding of the Sun's activity and behavior. This interdisciplinary approach highlights the importance of combining diverse data sources in solar physics research.

## 3. Method

This study focused on analyzing Doppler shift oscillations observed in BPs by the *IRIS*. The researchers used spectral lines sensitive to plasma temperature and velocity to construct

Doppler velocity time series. Wavelet analysis was performed to detect oscillation periods and understand the underlying physical processes. A deep learning model, based on a feedforward neural network, was trained to classify different oscillation patterns. The model utilized four components corresponding to different spectral lines observed by *IRIS*(see figure 1. Data pre-processing and segmentation were employed to handle the large volume of data. The accuracy of the model ranged from 75% to 99% in classifying oscillation patterns (see figure 2). The study provided insights into plasma behavior in the Sun's atmosphere and demonstrated the potential of deep machine learning in solar physics research (Sadeghi and Tavabi 2022a).

## 4. Results

In our study, we used the same methodological approach to analyze both short period damp and long period damp phenomena. The methods and techniques employed for data collec- tion, preprocessing, and analysis were consistent across both types of damping. However, despite using the same methodology, the results of our analysis revealed distinct differences between the two types of damping. These differences were observed in various aspects such as the characteristics of the damping behavior, the statistical properties, and the quantitative measurements associated with each type (tables 1 and 2).

**Table 1.** Short period damping parameters

|  | Net-QS | Inter-QS | Net-AR | Inter-AR | Net-CH | Inter-CH |
|---|---|---|---|---|---|---|
| Average max shift (Km/s) | 28 | 28 | 34 | 34 | 24 | 21 |
| Average damp (Seconds) | 216 | 120 | 131 | 220 | 150 | 121 |
| Average damp/period | 0.74 | 0.62 | 0.46 | 1.25 | 0.59 | 0.80 |
| Average delta | 0.50 | 0.39 | 0.88 | 0.73 | 0.63 | 0.45 |
| damp % | 62 | 49 | 32 | 21 | 70 | 55 |

*Notes*:

**Table 2.** average of long period damp components

| region | point type | f | $\tau$ | half-life | $\zeta$ | Q | repeat frequency |
|---|---|---|---|---|---|---|---|
| coronal hole | network | 0.8 m | 3500 | 2400 | 0.34 | 1.47 | 1500 |
|  | internetwork | 1.6 m | 2500 | 1700 | 0.13 | 3.84 | 1000 |
| active | network | 2.5 m | 2000 | 1400 | 0.33 | 1.48 | 4000 |
|  | internetwork | 1.7 m | 4800 | 3300 | 0.44 | 1.14 | 11000 |
| quiet | network | 1 m | 3000 | 2000 | 0.32 | 1.56 | 1200 |
|  | internetwork | 1.5 m | 1900 | 1300 | 0.17 | 2.94 | 1000 |

*Notes*:

## 5. Discussion

The study investigated the presence of damping in both short-period and long-period oscillations in the Sun's atmosphere.

Short-period damping:

• Previous studies suggested longitudinal waves as a mechanism for transporting energy in the Sun's corona but found their energy insufficient.

• The study analyzed bright points in active region (AR) and coronal hole (CH) regions (see figure 3).

• Internetwork regions in AR and CH displayed shorter decay times and half-lives compared to network regions, indicating higher damping rates potentially due to small-scale magnetic fields.

• CH regions exhibited shorter decay times and half-lives compared to AR regions, likely due to lower density and weaker magnetic fields.

• Bright points in both network and internetwork regions showed damped oscillations, with varying degrees of damping observed in different regions.

Long-period damping:

• Spectral analysis revealed the presence of longitudinal oscillations with longer periods in the Sun's transition region and corona.

• These oscillations exhibited outward-propagating Doppler motions, with amplitudes ranging from 10 to 35 km/s and periods of 180-550 s.

• The oscillations in both network and internetwork bright points were expected to have sufficient energy to drive the solar wind and heat the corona.

• The study did not specifically discuss the damping behavior of long-period oscillations.

Overall, the study provided insights into the damping behavior of both short-period and long-period oscillations in the Sun's atmosphere, contributing to our understanding of the dynamics and energy transport processes in the solar corona and solar wind.

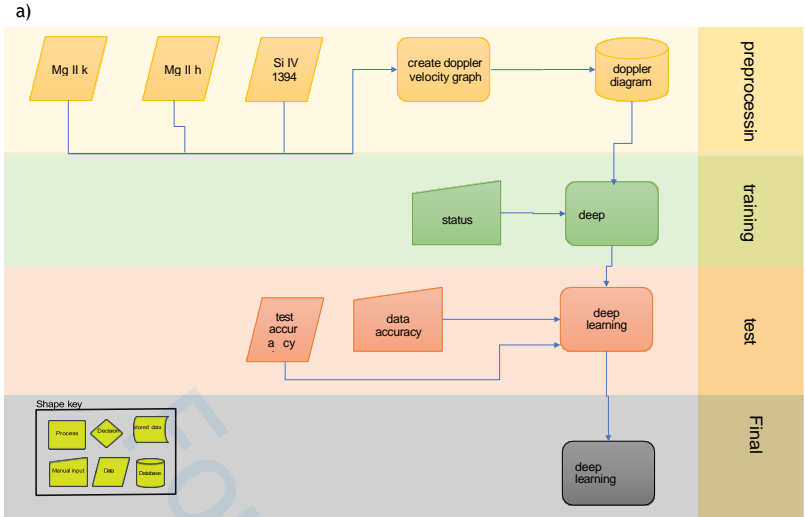

a)

b)

| | Name | type | Activation | Learnable | Total Learnable | States |
|---|---|---|---|---|---|---|
| 1 | sequenceinput | sequence input | 6 | - | 0 | Hidden State 500x1<br>Cell State 500x1 |
| 2 | lstm | LSTM | 500 | input weights 2000x6<br>recurrent weights 2000x500<br>Bias 2000x1 | 1014000 | - |
| 3 | fc | Fully connected | 6 | weights 6x500<br>Bias 6x1 | 3006 | - |
| 4 | softmax | Softmax | 6 | - | - | - |
| 5 | classoutput | Classification Output | 4 | - | - | - |

**Figure 1.** he deep machine learning examination of the finest time-series *IRIS* rasters validates the rhythmic regime of longitudinal waves discovered using Doppler maps. This research looked at the Doppler shift oscillations over time and above the *IRIS* bright spots. For each bright point, four-time series data were obtained and matched to the blue and red Doppler shifts of the Mg II h & k and Si IV spectra. The average period of Doppler velocity oscillations for network and internetwork points is 300 and 202 seconds, respectively, and bright points are now classified into six types: network bright points in an active region, internetwork bright points in an active region, network bright points in a quiet region, internetworks in a coronal hole area, and internetworks in a coronal hole region. This work employed 16 training data series to build a deep learning model with four components: Blue Si IV, Red Si IV, Blue Mg II, and Red Mg II. The model has 500 hidden layers and classifies data as moist, a single peak, rising, or none. The model's accuracy ranged from 58% to 98%, with an average of more than 80%. The model's validity was established by measuring two points in the active zone and two points in the coronal hole region.

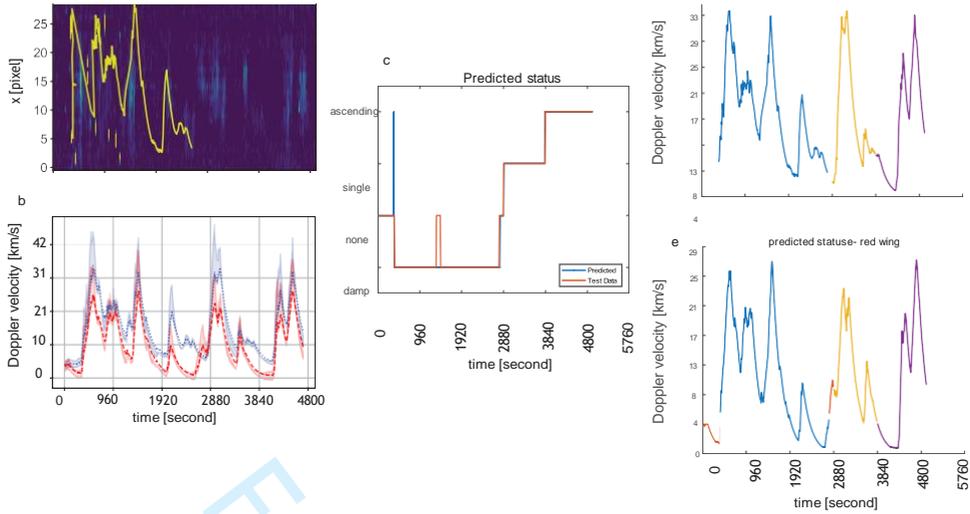

**Figure 2.** a) This figure illustrates the time series of the silicon 1394 $\mathring{A}$ spectrum. An indication of dampness can be observed in this image. b) The red and blue wings in this diagram depict the Doppler velocity shift. It can be noted that a blue shift corresponds to the occurrence of redshift dampening. c) In this graphic, the guide data and projected data are presented together. This information is related to the Si IV 1394 $\mathring{A}$ redshift. d, e) The predictions for two types of shifts, red and blue, are shown. Blue indicates attenuation, orange represents a single peak, and purple signifies an increasing trend.

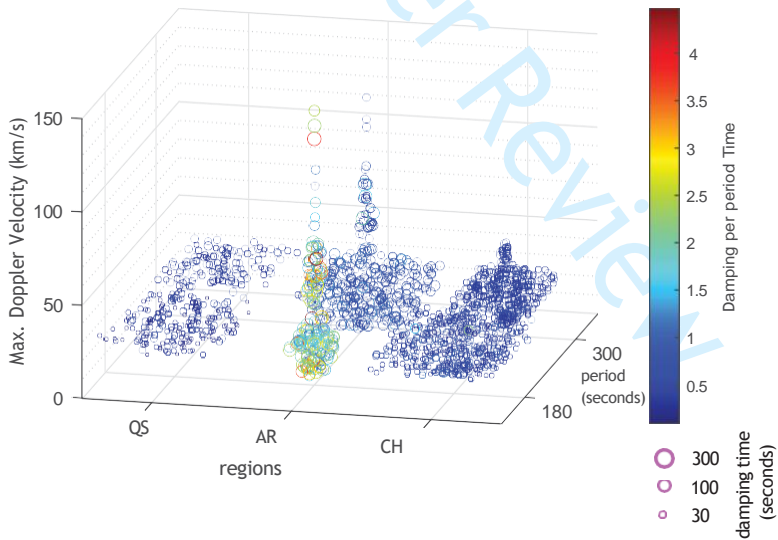

**Figure 3.** The swarm chart illustrates the maximum Doppler velocity for different regions (QS, AR, and CH) across two periods (300 and 180 sec.). The marker size represents the damping time, with larger markers indicating higher damping values. The chart showcases the distribution of data points and allows for visual comparison between regions and periods. The color bar indicates the corresponding damping per period time values. First, as we expected the number of net and Internetwork BPs numbers is much higher in CH, because the background is darker, also the Damp/Period ratio is the lowest at there, second, in AR we see a jump of Doppler velocity and very higher ratio of Damp/Period parameter at the net and internetwork BPs, the Damp/Period at the internetwork places is significantly growing up in some cases. Finally in the QS regions we could not see the meaningful differences between net and internetwork BPs.

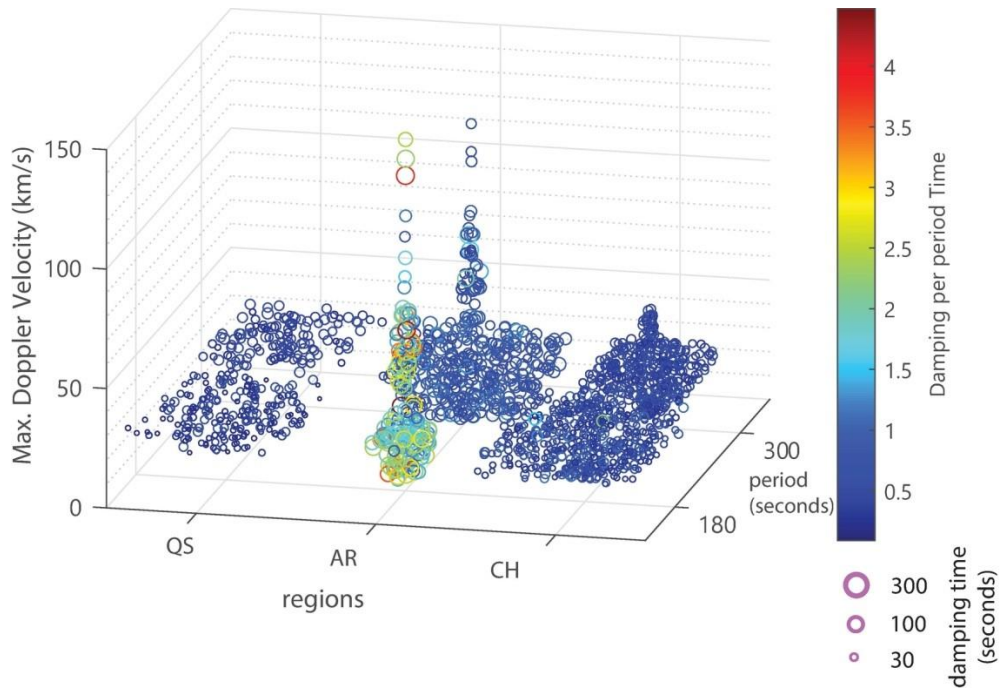

The swarm chart illustrates the maximum Doppler velocity for different regions (QS, AR, and CH) across two periods (300 and 180 sec.). The marker size represents the damping time, with larger markers indicating higher damping values. The chart showcases the distribution of data points and allows for visual comparison between regions and periods. The color bar indicates the corresponding damping per period time values. First, as we expected the number of net and Internetwork BPs numbers is much higher in CH, because the background is darker, also the Damp/Period ratio is the lowest at there, second, in AR we see a jump of Doppler velocity and very higher ratio of Damp/Period parameter at the net and internetwork BPs, the Damp/Period at the internetwork places is significantly growing up in some cases. Finally in the QS regions we could not see the meaningful differences between net and internetwork BPs.

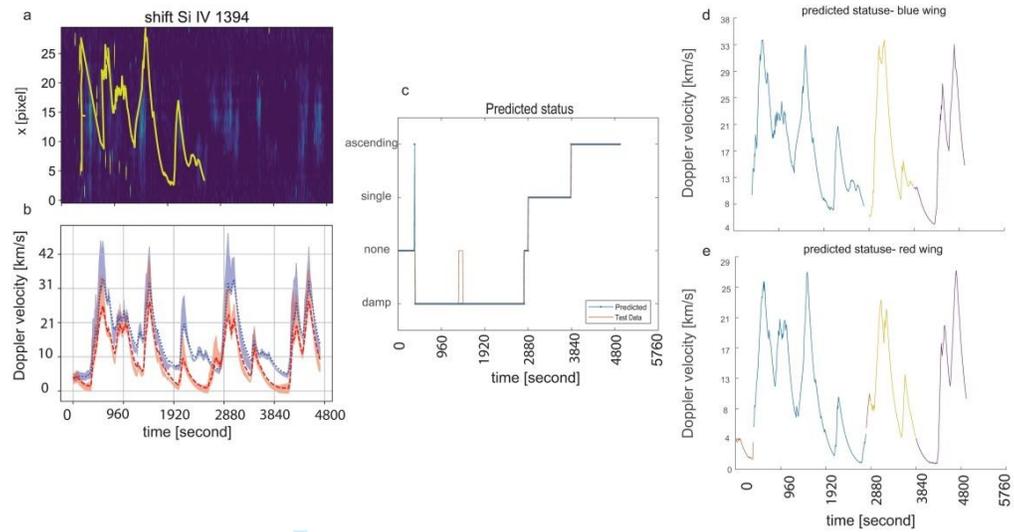

a) This figure illustrates the time series of the silicon 1394 $\dot{A}$ spectrum. An indication of dampness can be observed in this image.

b) The red and blue wings in this diagram depict the Doppler velocity shift. It can be noted that a blue shift corresponds to the occurrence of redshift dampening.

c) In this graphic, the guide data and projected data are presented together. This information is related to the Si {\footnotesize IV} 1394 $\dot{A}$ redshift.

d, e) The predictions for two types of shifts, red and blue, are shown. Blue indicates attenuation, orange represents a single peak, and purple signifies an increasing trend.

a)

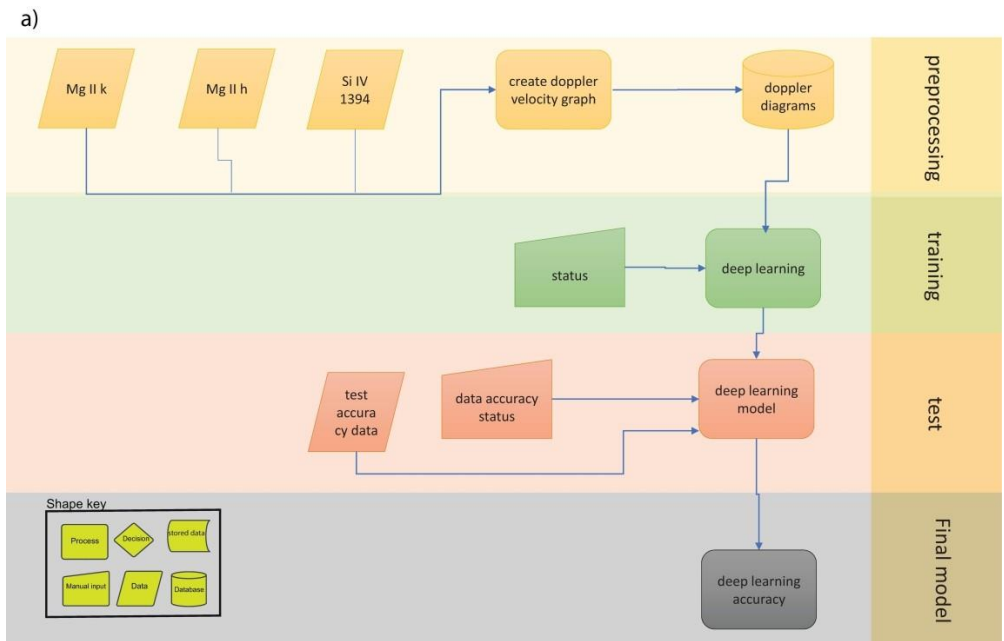

b)

|   | Name | type | Activation | Learnable | Total Learnable | States |
|---|------|------|------------|-----------|-----------------|--------|
| 1 | sequenceinput | sequence input | 6 | - | 0 | Hidden State 500x1 Cell State 500x1 |
| 2 | lstm | LSTM | 500 | input weights 2000x6 recurrent weights 2000x500 Bias 2000x1 | 1014000 | - |
| 3 | fc | Fully connected | 6 | weights 6x500 Bias 6x1 | 3006 | - |
| 4 | softmax | Softmax | 6 | - | - | - |
| 5 | classoutput | Classification Output | 4 | - | - | - |

The deep machine learning examination of the finest time-series \emph{IRIS} rasters validates the rhythmic regime of longitudinal waves discovered using Doppler maps. This research looked at the Doppler shift oscillations over time and above the \emph{IRIS} bright spots. For each bright point, four-time series data were obtained and matched to the blue and red Doppler shifts of the Mg {\footnotesize II h \& k} and Si {\footnotesize IV} spectra. The average period of Doppler velocity oscillations for network and internetwork points is 300 and 202 seconds, respectively, and bright points are now classified into six types: network bright points in an active region, internetwork bright points in an active region, network bright points in a quiet region, internetworks in a coronal hole area, and internetworks in a coronal hole region. This work employed 16 training data series to build a deep learning model with four components: Blue Si {\footnotesize IV}, Red Si {\footnotesize IV}, Blue Mg {\footnotesize II}, and Red Mg {\footnotesize II}. The model has 500 hidden layers and classifies data as moist, a single peak, rising, or none. The model's accuracy ranged from 58\% to 98\%, with an average of more than 80\%. The model's validity was established by measuring two points in the active zone and two points in the coronal hole region.